\documentclass{emulateapj}

\usepackage{apjfonts}
\usepackage{amssymb}
\usepackage{epsfig}
\usepackage{amsmath}
\usepackage{graphicx} 
\usepackage{relsize}
\usepackage[toc,page]{appendix}
\usepackage[T1]{fontenc}
\usepackage{chngcntr}

\usepackage[backref,breaklinks,colorlinks,citecolor=blue]{hyperref}
\usepackage[all]{hypcap}

\usepackage[caption = false]{subfig}

\begin{document}

 \title{Quasar Feedback in the Ultraluminous Infrared Galaxy
   F11119$+$3257: Connecting the Accretion Disk Wind with the
   Large-Scale Molecular Outflow}

\author{S. Veilleux\altaffilmark{1,2}, A. Bolatto\altaffilmark{1,2},
  F. Tombesi\altaffilmark{3,4,5}, M. Mel\'endez\altaffilmark{1,6,7},
  E. Sturm\altaffilmark{8}, E. Gonz\'alez-Alfonso\altaffilmark{9},
  J. Fischer\altaffilmark{10}, \& D. S. N. Rupke\altaffilmark{11}}

\altaffiltext{1}{Department of Astronomy, University of Maryland,
  College Park, MD 20742, USA; veilleux@astro.umd.edu}

\altaffiltext{2}{Joint Space-Science Institute, University of Maryland,
  College Park, MD 20742, USA}

\altaffiltext{3}{X-ray Astrophysics Laboratory, NASA Goddard Space
  Flight Center, Greenbelt, MD 20771, USA}

\altaffiltext{4}{Department of Astronomy and CRESST, University of
  Maryland, College Park, MD 20742, USA}

\altaffiltext{5}{Dipartimento di Fisica, Universit\`{a} di Roma Tor
  Vergata, Via della Ricerca Scientifica 1, I-00133 Roma, Italy}

\altaffiltext{6}{NASA Goddard Space Flight Center, Greenbelt, MD
  20771, USA}

\altaffiltext{7}{Wyle Science, Technology and Engineering Group, 1290
  Hercules Avenue, Houston, TX 77058 USA}

\altaffiltext{8}{Max-Planck-Institute for Extraterrestrial Physics
  (MPE), Giessenbachstrasse 1, D-85748, Garching, Germany}

\altaffiltext{9}{Departamento de F\'isica y Matem\'aticas, Universidad
  de Alcal\'a, Campus Universitario, E-28871 Alcal\'a de Henares,
  Madrid, Spain}

\altaffiltext{10}{Naval Research Laboratory, Remote Sensing Division,
  4555 Overlook Avenue SW, Washington, DC 20375, USA}

\altaffiltext{11}{Department of Physics, Rhodes College, Memphis, TN
  38112, USA}

\begin{abstract}
In Tombesi et al.\ (2015), we reported the first direct evidence for a
quasar accretion disk wind driving a massive molecular outflow. The
target was F11119$+$3257, an ultraluminous infrared galaxy (ULIRG)
with unambiguous type-1 quasar optical broad emission lines. The
energetics of the accretion disk wind and molecular outflow were found
to be consistent with the predictions of quasar feedback models where
the molecular outflow is driven by a hot energy-conserving bubble
inflated by the inner quasar accretion disk wind.  However, this
conclusion was uncertain because the energetics of the outflowing
molecular gas were estimated from the optically thick OH 119 $\mu$m
transition profile observed with {\em Herschel}. Here, we
independently confirm the presence of the molecular outflow in
F11119$+$3257, based on the detection of broad wings in the CO(1$-$0)
profile derived from ALMA observations. The broad CO(1$-$0) line
emission appears to be spatially extended on a scale of at least
$\sim$7 kpc from the center. Mass outflow rate, momentum flux, and
mechanical power of (80 $-$ 200) $R_7^{-1}$ $M_\odot$ yr$^{-1}$, (1.5
$-$ 3.0) $R_7^{-1}$ $L_{\rm AGN}/c$, and (0.15 $-$ 0.40)\% $R_7^{-1}$
$L_{\rm AGN}$ are inferred from these data, assuming a CO$-$to$-$H$_2$
conversion factor appropriate for a ULIRG ($R_7$ is the radius of the
outflow normalized to 7 kpc and $L_{\rm AGN}$ is the AGN luminosity).
These rates are time-averaged over a flow time scale of $7 \times
10^6$ yrs. They are similar to the OH-based rates time-averaged over a
flow time scale of $4 \times$ $10^5$ yrs, but about a factor 4 smaller
than the local (``instantaneous''; $\la$ $10^5$ yrs) OH-based
estimates cited in Tombesi et al. The implications of these new
results are discussed in the context of time-variable quasar-mode
feedback and galaxy evolution. The need for an energy-conserving
bubble to explain the molecular outflow is also re-examined.
\end{abstract}


\section{Introduction}

Rapidly accreting supermassive black holes (SMBHs) produce tremendous
amounts of radiative energy. The coupling of this energy with gas near
the black hole or at larger scales in the host galaxy produces
outflows of gas. These ``quasar-mode'' outflows are distinct from
``radio-mode'' jets in that they are much less collimated, and
therefore have the potential to impact a much greater swath of a
galaxy's gas.  Quasar-mode outflows are often invoked to play a
fundamental role in the evolution of both SMBHs and their host
galaxies, quenching star formation and explaining the tight
SMBH-galaxy relations (e.g., Veilleux et al.\ 2005; Fabian 2012).
Recent observations of large-scale neutral and molecular outflows in
(U)LIRGs have provided supporting evidence for this idea, as they
directly trace the gas out of which stars form (e.g., Fischer et
al.\ 2010; Feruglio et al.\ 2010, 2015; Sturm et al.\ 2011; Alatalo et
al.\ 2011, 2015; Rupke \& Veilleux 2011, 2013a, 2013b, 2015; Veilleux
et al.\ 2013, hereafter V13; Aalto et al.\ 2012, 2015; Cicone et
al.\ 2014; Garc\'ia-Burillo et al.\ 2015; Lindberg et al.\ 2016;
Gonz\'alez-Alfonso et al.\ 2014, 2017). Theoretical models suggest an
origin of these outflows as energy-conserving flows driven by fast AGN
accretion disk winds (e.g., Zubovas \& King 2012, 2014;
Faucher-Gigu\`ere \& Quataert 2012; Costa et al.\ 2014; Nims et
al.\ 2015).

Our previous claims of a connection between large-scale molecular
outflows and AGN activity in (U)LIRGs were based on the fact that
systems with quasar-like AGN luminosities host the faster and more
powerful outflows (V13; Rupke \& Veilleux 2013a; Cicone et al.\ 2014).
Until recently, these claims were incomplete because they were lacking
the detection of the putative inner wind. Conversely, studies of
powerful AGN accretion disk winds to date had focused only on X-ray
observations of local radio-quiet and radio-loud AGN and a few higher
redshift quasars, but had ignored the impact of these winds on the
galaxy host (e.g., Tombesi et al.\ 2010, 2014; Nardini et al.\ 2015
and references therein). 

This situation changed with the publication of Tombesi et al.\ (2015,
hereafter T15), where we showed the clear (6.5-$\sigma$) detection of
a powerful AGN accretion disk wind with a mildly relativistic velocity
of $\sim$0.25 $c$ in the X-ray spectrum of IRAS F11119$+$3257, a
nearby ($z$ = 0.190; 1$^{\prime\prime}$ = 3.19 kpc) optically
classified type 1 ULIRG hosting a powerful molecular outflow with
velocity $V_{\rm out,OH}$ = 1000 $\pm$ 200 km~s$^{-1}$. This was the
first {\em direct} evidence for a fast quasar accretion disk wind
driving a large-scale molecular outflow.$\footnote{Since the
  publication of T15, Feruglio et al.\ (2015) has reported the
  tentative detection of a $\sim$0.1 $c$ X-ray wind in Mrk~231 at the
  3.5-$\sigma$ level (see also Reynolds et al.\ 2017).}$ The
energetics of the accretion disk wind and molecular outflow derived
from our data are consistent with the energy-conserving mechanism
(T15). In this scenario, the violent interaction of the fast
inner AGN wind with the ISM of the host results in shocked wind gas
that does {\em not} efficiently cool, but instead expands
adiabatically as a hot bubble (e.g., Zubovas \& King 2012, 2014;
Faucher-Gigu\`ere \& Quataert 2012; Costa et al.\ 2014; Nims et
al.\ 2015). The adiabatically expanding shocked wind sweeps up gas and
drives an outer shock into the host ISM. The outflowing gas cools
radiatively, and most of it ``freezes out'' into clumps of cold
molecular material. This picture is also able to explain the existence
of a fast ($\sim$1300 km~s$^{-1}$) neutral-atomic (Na~I~D) outflow in
this system (Rupke et al.\ 2005b).  A variant on this scenario is that
pre-existing molecular clouds from the host ISM are entrained in the
adiabatically expanding shocked wind, accelerated to the observed
velocities without being destroyed by the many erosive forces and
instabilities (e.g., Cooper et al.\ 2009; McCourt et al.\ 2015, 2016;
Scannapieco \& Br\"uggen 2015; Banda-Barrag\'an et al.\ 2016; Tanner,
Cecil, \& Heitsch 2016; Thompson et al.\ 2015, 2016; Scannapieco 2017)

While the existence of the molecular outflow in F11119$+$3257 is
unquestionable based on the OH absorption profile, the energetics of
this outflow remain uncertain. In T15, we cite a mass
outflow rate $\dot{M}_{\rm out,OH}$ = 800$^{+1200}_{-550}$ $M_\odot$
yr$^{-1}$, a momentum flux log $\dot{P}_{\rm out,OH}$ = 36.7 $\pm$ 0.5
in dyne, and a mechanical power log $\dot{E}_{\rm out,OH}$ = 44.4
$\pm$ 0.5 in erg s$^{-1}$.  The large uncertainties stem mainly from
the high optical depth of the OH 119 $\mu$m line and the lack of
higher excitation line profiles (e.g., OH 65 and 84 $\mu$m), and to
some degree from the fact that the OH outflow is not spatially
resolved in the {\em Herschel} data. In T15, we had to compare the
predictions of our radiative transfer models (e.g., Gonz\'alez-Alfonso
et al.\ 2014) with the observed velocity-resolved profile of OH 119
$\mu$m to constrain the location (0.1 -- 1.0 kpc)
of the OH molecules that produce the OH profile. The energetics of the
OH outflow scale linearly with the OH abundance $X_{\rm OH}$ =
OH/H.$\footnote{Note, however, that the OH abundance adopted in T15,
  $X_{\rm OH}$ = 2.5 $\times$ 10$^{-6}$, is, within a factor of
  $\sim$3, consistent with the value inferred in the Galactic Sgr B2,
  the Orion KL outflow, and in buried galaxy nuclei, as well as with
  the predictions of chemical models of dense photodissociation
  regions and of cosmic-ray and X-ray dominated regions (see
  Gonz\'alez-Alfonso et al.\ 2017).}$ It is also important to note
that the energetics cited in T15 are the local (``instantaneous'')
quantities estimated at a radius $R_{\rm out,OH}$ = 300 pc. These
values are time-averaged over $\Delta R_{\rm out,OH}/V_{\rm out,OH}$
$\la$ $10^5$ yrs, the time the outflowing shell of material takes to
cross the shell thickness $\Delta R_{\rm out,OH}$ $\sim$ 75 pc. The
values time-averaged over the flow time scale $R_{\rm out,OH}/V_{\rm
  out,OH}$ are smaller by a factor of 4.

In the present paper, we take a complementary approach to constrain
the energetics of the molecular outflow in F11119$+$3257, using line
emission from low-level transitions of CO as a tracer of the
outflowing molecular material. IRAM 30-m observations by Xia et
al.\ 2012 have already shown that the CO(1$-$0) emission in
F11119$+$3257 is broad (FWHM $\sim$ 285 $\pm$ 36 km s$^{-1}$),
centered on redshift $z_{CO} = 0.190$, and has a luminosity of 1.12
$\times$ 10$^{10}$ K km s$^{-1}$ pc$^2$, corresponding to a molecular
gas mass of $\sim$9 $\times$ 10$^9$ M$_\odot$ for a Galactic
conversion factor of $\alpha_{CO}$ = 4.3 $M_\odot$ (K km s$^{-1}$
pc$^2$)$^{-1}$. Our new ALMA data are considerably more sensitive than
the IRAM data and reveal faint broad wings in the CO(1$-$0) line
emission profile.  Section 2 describes the ALMA observations of
F11119$+$3257. The results from the analysis of these data are
presented in Section 3 and compared with those of T15 in Section
4. The main conclusions are summarized in Section 5.  Throughout this
paper, we adopt a redshift $z = 0.190$ (Rupke et al.\ 2005b, 2017 in
prep.; Xia et al.\ 2012), a luminosity distance of 933 Mpc,
and corresponding linear scale of 3.19 kpc per arcsecond for
F11119$+$3257 (using $H_0$ = 69.6 km~s$^{-1}$~Mpc$^{-1}$, $\Omega_M$ =
0.286, and $\Omega_\Lambda$ = 0.714 from Bennett et al.\ 2014).

\section{ALMA Observations and Data Reduction}

F11119$+$3257 was observed for 1.9 hours (on-source, 3.2 hours in
total) on 2016 January 3 and 10 as part of project 2015.1.00305.S in
Cycle 3. The observations were carried out with 36 antennas in the
compact C36-1 12-m array configuration with baselines 15 -- 312 m,
resulting in an angular resolution of $\sim$2$\farcs$8 $\sim$ 9
kpc. The main objective of these observations was to detect the
molecular outflow based on the presence of broad wings in the
CO(1$-$0) line emission at 96.9 GHz (Band 3).  A RMS (1-sigma)
sensitivity of 140 $\mu$Jy in each 100 km s$^{-1}$ channel was
targeted, corresponding to 0.7\% of the peak flux density measured by
Xia et al.\ 2012 (4 mK $\sim$ 20 mJy). The CO(1$-$0) wing-to-peak
ratios in (U)LIRGs (e.g., Cicone et al.\ 2014) are typically
$\sim$5\%, or 1.0 mJy $\sim$ 7 $\sigma$ for F11119$+$3257. A similar
result is obtained if one assumes that the CO outflow luminosity is
roughly proportional to the OH outflowing mass. For this exercise, we
use a OH-to-CO scaling factor to translate the OH 119 $\mu$m
equivalent width into CO(1$-$0) line flux based on the average
observed relation in the outflows of ULIRGs Mrk~231, IRAS
F08572$+$3915, and IRAS F10565$+$2448 (the scatter in the relation is
$\sim$30\%; Fig.\ 2 in Gonz\'alez-Alfonso et al.\ 2017). Only the blue
wing ($\le$ $-$200 km/s) of the OH line equivalent width (i.e.\ only
the truly outflowing component) is used for this calculation.

The requested angular resolution ($\sim$3$\arcsec \sim$ 9 -- 10 kpc)
and largest angular scale ($\sim$15 $-$ 20$\arcsec$) of these objects
safely allow complete CO(1$-$0) flux recovery from this galaxy (broad
wings + bright core emission near systemic velocity; Xia et
al.\ 2012).  The correlator setup was optimized to simultaneously
observe CO(1$-$0) and the adjacent continuum emission. The central
frequency of baseband-4 was adjusted so that baseband-4 was contiguous
with baseband-3 and also covered ``for free'' the CN (1-0) complex at
113.15 and 113.50 GHz, and the SiO $v$=0 (3-2), $v=1$ (3-2), and $v=2$
(3-2) transitions at 130.269, 129.363, and 128.459 GHz, respectively,
possible tracers of shocked molecular gas in this galaxy. Spectral
smoothing by a factor of 4 was used to reduce the data rate while
maintaining a reasonable velocity resolution of $\sim$ 6
km~s$^{-1}$. The pipeline-calibrated interferometric visibilities
delivered by ALMA were continuum-subtracted in the {\em uv}-plane using a
first-order polynomial, then imaged at 20 km~s$^{-1}$ resolution using
Briggs weighting with a robust parameter of 0.5 and cleaned using a
tight box around the source. The restoring beam is 3$\farcs$46
$\times$ 2$\farcs$21 FWHM with PA = 12$^\circ$.

\clearpage

\section{Results}

Figure 1 shows the full continuum-subtracted upper sideband (USB)
spectrum extracted from within a circular 3\arcsec-radius aperture
centered on the source.  The channels are 20 km~s$^{-1}$ wide but
Hanning velocity smoothing was carried out to result in a spectral
resolution of $\sim$40 km~s$^{-1}$. The cube has a noise of 0.28
  mJy in the 20 km~s$^{-1}$ channels. The strong CO(1$-$0) line
emission is detected with a signal-to-noise ratio (S/N) of
  $\sim$65 at the peak. The hyperfine components of CN (1-0) at
113.15 and 113.50 GHz are also detected. On the other hand, the SiO
$v$=0, 1, 2 (3-2) transitions at 130.269, 129.363, and 128.459 GHz are
not visible in our band-4 data, so they are not discussed any further
in the remainder of the paper.

Figure 2 zooms in on the CO(1$-$0) line emission within circular
apertures with radii of 3\arcsec\ and 5\arcsec\ centered on
F11119$+$3257. Broad line emission indicative of an outflow is
detected in both panels out to velocities $\sim$ $\pm$1000 km~s$^{-1}$
relative to systemic ($z$ = 0.190), remarkably similar to the velocity
of the OH outflow reported in T15. Three methods are used to quantify
the strength of this broad emission.

First, we carry out a simultaneous fit for two Gaussians (one narrow,
one broad) to these data. The broad Gaussian is shown as the yellow
area in Figure 2. The residuals are generally less than $\pm$ 0.5
mJy. The quantities derived from these fits are listed in Table 1. The
uncertainties on these quantities were estimated using a bootstrap
Monte Carlo method.  Note that the broad-to-narrow peak flux ratios
($\sim$3$-$4\%) listed in that table are similar to those assumed for
the requested ALMA time ($\sim$5\%; Section 2). The fact that the
broad-to-narrow integrated flux ratio is larger in the 5\arcsec-radius
aperture spectrum (0.19) than in the 3\arcsec-radius aperture spectrum
(0.14) suggests that the broad line emission extends out to a radius
of 5\arcsec, although the 5\arcsec\ spectrum is noticeably more noisy
than the 3\arcsec\ spectrum. The fluxes of the broad components
derived from these fits are considered upper limits to the actual flux
from the outflowing material since they include CO line emission at
low velocities which may not be associated with the outflow. We
attempt to remove this low-velocity material using a different
strategy.

Figure 3 reproduces the continuum-subtracted CO (1$-$0) spectrum
integrated over a circular aperture with a radius of 3$\arcsec$.  The
red line shows the original spectrum (cut off vertically to show the
details in the wings of CO(1$-$0)).  The blue line shows the residuals
after fitting and removing a Gaussian source model to each 20
km~s$^{-1}$ channel. First, a two-dimensional Gaussian was fit to the
image for each channel in the region where a source is detected. The
results were then used to make a smooth source model with linearly
changing position as a function of velocity (to account for a possible
velocity gradient; see below), Gaussian changing intensity, but
constant size and orientation. This smooth source model was then
removed from each velocity slice to arrive at a ``residuals'' cube.
The high S/N of the detection allows us to centroid the source in each
channel with very good sub-beam precision. The velocity gradient
measured is +350 km~s$^{-1}$ kpc$^{-1}$ in right ascension and $-$200
km~s$^{-1}$ kpc$^{-1}$ in declination (Figure 4). This compares well
with the direction and amplitude of the velocity gradient measured in
an unpublished Keck laser guide star adaptive optics Pa$\alpha$ data
cube of F11119$+$3257 obtained with OSIRIS (D.\ S.\ N.\ Rupke 2017,
private communication).  Assuming that this represents the rotation of
the gas in the host galaxy, a dynamical mass of $\sim$5 $\times$
10$^9$ $M_\odot$ within $\sim$1 kpc from the center is derived from
these data.

The yellow region in Figure 3 shows the CO ``high velocity'' emission,
which cannot be accounted for by the gas in pure rotation.  Figure 5
shows maps of the rotating material and high-velocity gas integrated
over the ``residuals'' channels shown in yellow in Figure 3. Table 2
lists the parameters derived from Figure 5c. The high-velocity gas is
extended and offset by $+$0$\farcs$22 $\pm$ 0$\farcs$05 in right
ascension and $-$0$\farcs$75 $\pm$ 0$\farcs$10 in declination from the
USB band-4 continuum emission (shown in Fig.\ 1b). A Gaussian fit to
the high-velocity line emission image of Figure 3 finds a FWHM size of
5$\farcs$1 $\times$ 2$\farcs$8 with 0$\farcs$1 uncertainty in either
direction, elongated along PA = 4$^\circ$ from the North-South
direction; this is significantly larger than the 3$\farcs$46 $\times$
2$\farcs$21 FWHM beam.

Aperture photometry on the high-velocity gas confirms that it is
indeed extended. In Figure 6, the enclosed high-velocity integrated
flux peaks around a radius of 5$\arcsec$ $\pm$ 1$\farcs$5 and then
remains roughly constant around $\sim$0.4 $\pm$ 0.1 Jy km
s$^{-1}$. This is our conservative estimate for the flux from the
molecular outflow. The uncertainty on the enclosed flux is estimated
from the amplitude of the fluctuations around the value of 0.4 Jy km
s$^{-1}$ observed in Figure 6. A radius of 5$\arcsec$ $\pm$ 1$\farcs$5
on the image corresponds to an actual radius of
4$\farcs$8$^{+1.5}_{-2.7}$, after correcting for the beam size
(3$\farcs$46 $\times$ 2$\farcs$21 FWHM i.e.\ $\sim$2$\farcs$8
FWHM). This radius, 4$\farcs$8$^{+1.5}_{-2.7}$ = 15$^{+5}_{-8}$ kpc,
is our best estimate of the maximum extent of the CO outflow.

As an independent check on the results from our analysis of these
  imaging data, we also derived the sizes and fluxes of the wing
  emission by fitting the data directly in the {\em uv} plane. For
  this exercise, we used both the CASA {\em uv}-Plane Model Fitting
  routine {\em uvmodelfit} and {\em uvmultifit}, the library of
  Mart\'i-Vidal et al.\ (2014). The results are summarized in Table
  3. In contrast to {\em uvmodelfit}, {\em uvmultifit} could not deal
  with the sum of the red and blue wings, so they were fit
  separately. The signals were integrated between 96.9893 and 97.1315
  GHz ($-$820 to $-$380 km s$^{-1}$) for the blue wing, and 96.6081 to
  96.7244 GHz for the red wing ($+$440 to $+$800 km s$^{-1}$). As
  shown in Figure 3, these channels are not affected by rotation so we
  did not have to remove a disk model in the {\em uv}-plane
  fitting. The results from the two {\em uv} fitters are consistent
  with each other and with the results from the imaging methods
  (compare the entries in Table 3 with those of Tables 1 and 2).

Taken at face value and keeping in mind the large uncertainties
  on these estimates, the CO outflow in F11119$+$3257 is the largest
  molecular outflow so far detected in a local ULIRG: $R_{\rm
  out,CO}$ is typically $\sim$1 kpc in these objects with the possible
exceptions of F23060$+$0505 and Mrk~876, where the CO outflows are not
well resolved and imply $R_{\rm out,CO}$ $\le$ 4.05 kpc and $\le$ 3.55
kpc, respectively (Cicone et al.\ 2014), and F08572$+$3915, where a
fast-moving ($\sim$1000 km~s$^{-1}$) cloud was recently detected by
Janssen et al.\ (2017, in prep.) at $\sim$6 kpc from the NW
galaxy. Nothing in the galaxy host of F11119$+$3257 itself (Kim et
al.\ 2002; Veilleux et al.\ 2002) can account for the morphology and
kinematics of this broad CO line emission. The implications of these
results are discussed in the following section.

\section{Discussion}

\subsection{Energetics of the CO Outflow}

The mass of molecular gas involved in the outflow can be derived from
the integrated flux densities quoted in the previous section (0.3 $-$
1.0 Jy km s$^{-1}$), using equation 3 from Bolatto, Wolfire, \& Leroy
(2013a). A Galactic CO$-$to$-$H$_2$ conversion factor $X_{\rm CO}$ = 2
$\times$ 10$^{20}$ cm$^{-2}$ (K km s$^{-1}$)$^{-1}$ (or equivalently
$\alpha_{\rm CO}$ = 4.3 $M_\odot$ (K km s$^{-1}$ pc$^2$)$^{-1}$) would
imply a CO-based molecular mass $M_{\rm out,CO}$ = (3 $-$ 8) $\times$
10$^9$ $M_\odot$, given a luminosity distance of 933 Mpc.  A
conservative lower limit on the outflowing molecular gas mass $M_{\rm
  out,CO}$ $\sim$ (2 $-$ 6) $\times$ 10$^8$ $M_\odot$ is derived if we
use the $\sim$13 $\times$ smaller optically thin $X_{\rm CO}$ used by
Bolatto et al.\ (2013b) to estimate the outflowing molecular gas mass
in NGC~253.  A compromise between these two extremes is to adopt a
ULIRG-like $\alpha_{\rm CO}$ of 0.8 $M_\odot$ (K km s$^{-1}$
pc$^2$)$^{-1}$ as done by Cicone et al.\ (2014). This results in an
outflowing molecular gas mass of $\sim$ (0.6 $-$ 1.4) $\times$ 10$^9$
$M_\odot$, which falls at the high mass end of the spectrum covered by
local ULIRGs (Cicone et al.\ 2014; Gonz\'alez-Alfonso et al.\ 2017).
For comparison, the non-outflowing material emits 5 -- 6 Jy km
s$^{-1}$ in CO(1$-$0).  Assuming the same ULIRG-like $X_{\rm CO}$ as
for the outflowing material, the amount of quiescent molecular gas in
the host galaxy is $M_{\rm host,CO}$ $\sim$ (7 $-$ 10) $\times$
10$^{9}$ $M_\odot$, i.e.\ in the top quartile of local ULIRGs and
infrared quasars (e.g., Solomon et al.\ 1997; Evans et al.\ 2001,
2006; Scoville et al.\ 2003; Xia et al.\ 2012), and about 5 $-$ 15
$\times$ the amount in the CO outflow.

The next step is to derive the CO-based mass outflow rate
$\dot{M}_{\rm out,CO}$, momentum flux $\dot{P}_{\rm out,CO}$, and
mechanical power $\dot{E}_{\rm out,CO}$ of the molecular outflow of
F11119$+$3257. As discussed in detail in Rupke et al.\ (2005a) and
Gonz\'alez-Alfonso et al.\ (2017), there are two limiting approaches
to the estimation of the outflow energetics: the local or
instantaneous (maximum) values and the average (minimum) values. The
local or instantaneous values are time-averaged over the time scale
taken by the outflow shell of material to cross the thickness of the
shell. This first approach was used in Sturm et al.\ (2011),
Gonz\'alez-Alfonso et al.\ (2014), and T15. Here, we use instead the
most conservative $\dot{M}_{\rm out}$, $\dot{P}_{\rm out}$, and
$\dot{E}_{\rm out}$ values based on the second approach to
characterize the outflow of F11119$+$3257. In that case, we have
\begin{eqnarray}
\dot{M}_{\rm out,CO} & = & \frac{M_{\rm out,CO}~V_{\rm out,CO}}{R_{\rm out,CO}}\\
\dot{P}_{\rm out,CO} & = & \dot{M}_{\rm out,CO}~V_{\rm out,CO}\\
\dot{E}_{\rm out,CO} & = & \frac{1}{2} \dot{M}_{\rm out,CO}~V_{\rm out,CO}^2.
\end{eqnarray}

These correspond to the ``time-averaged thin shell'' values of Rupke
et al.\ (2005a), time-averaged over the flow time scale $R_{\rm
  out,CO}/V_{\rm out,CO}$, and have been used extensively to describe
the energetics of the ionized and neutral phases of outflows (e.g.,
Rupke \& Veilleux 2013a; Arav et al.\ 2013; Borguet et al.\ 2013;
Heckman et al.\ 2015) as well as some molecular outflows
(Gonz\'alez-Alfonso et al.\ 2017).  They are most appropriate for
comparison with outflow models (e.g., Faucher-Gigu\`ere \& Quataert
2012; Thompson et al.\ 2015; Stern et al.\ 2016). In some studies
(e.g., Feruglio et al.\ 2010, 2015; Maiolino et al.\ 2012; Rodr\'iguez
Zaur\'in et al.\ 2013; Cicone et al.\ 2014; Harrison et al.\ 2014;
Garc\'ia-Burillo et al.\ 2015), a factor of 3 higher values have been
used under the assumption that the emitting spherical (or
multiconical) volume is filled with uniform density. However, for a
steady mass-conserving flow with constant velocity, we would expect a
density at the outer radius only 1/3 that of the average, thus also
yielding the expression in eq.\ (1).  The quick drop-off in the radial
intensity profile of the outflow emission of F11119$+$3257 indeed
seems consistent with this picture. Thus, we adopt eqs. (1)-(2)-(3)
for the rest of the discussion.

Our choice of $R_{\rm out,CO}$ and $V_{\rm out,CO}$ will set the flow
time scale ($R_{\rm out,CO}/V_{\rm out,CO}$) in these expressions and
therefore has to be done with care. If, for instance, we set $R_{\rm
  out,CO}/V_{\rm out,CO}$ = $R_{\rm out,CO}^{\rm max}/V_{\rm
  out,CO}^{\rm max}$, where $R_{\rm out,CO}^{\rm max}$ is the maximum
extent of the outflow ($\sim$15 kpc; Sec.\ 3) and $V_{\rm out,CO}^{\rm
  max}$ is the maximum outflow velocity ($\sim$1000 km s$^{-1}$,
ignoring projection effects), then the flow time scale $R_{\rm
  out,CO}/V_{\rm out,CO}$ $\approx$ 1.5 $\times$ 10$^7$ yrs. This
value would underestimate the actual flow time if all of the
outflowing gas originated from the center and was uniformly
accelerated from rest. If this were the case, the flow time scale
would be longer by a factor $\sim$ 2 (Gonz\'alez-Alfonso et al.\ 2017)
and we would expect the material with the highest outflow velocities
to be located further from the center than the material with the
lowest outflow velocities. While our data are not deep enough to allow
us to detect any velocity gradient in the outflow emission (Fig.\ 5),
systematic positive radial velocity gradients have not been detected
in the data of any other ULIRG (e.g., Cicone et al.\ 2014). Thus, we
do not favor this longer flow time scale.

We argue instead for a smaller flow time scale given that our
measurement of the full extent of the outflowing gas is uncertain and
a significant fraction ($\ga$50\%) of the outflowing material is
unresolved (Fig.\ 5). If the outflow were in fact completely
unresolved, $R_{\rm out,CO}$ $\la$ 4 kpc and $R_{\rm out,CO}/V_{\rm
  out,CO}$ $\la$ 4 $\times$ 10$^6$ yrs. The actual value of ($R_{\rm
  out,CO}/V_{\rm out,CO}$) most likely lies between these two
extremes.  In the following discussion, we adopt a conservatively low
value for the radius, $R_{\rm out,CO}$ = 7 kpc, and $V_{\rm out,CO}$ =
1000 km s$^{-1}$ (hence $R_{\rm out,CO}/V_{\rm out,CO}$ = 7 $\times$
10$^6$ yrs), as nominal values of the size and velocity of the CO
outflow, and a ULIRG-like CO$-$to$-$H$_2$ conversion factor (we
discuss the validity of this latter assumption in Section 4.3). From
eqs.\ (1) $-$ (3), we get $\dot{M}_{\rm out,CO}$ = (80 $-$ 200)
$M_\odot$ yr$^{-1}$, $\dot{P}_{\rm out,CO}$ = (6 $-$ 13) $\times$
10$^{35}$ dyne = (1.5 $-$ 3.0) $L_{\rm AGN}/c$, and $\dot{E}_{\rm
  out,CO}$ = (3 $-$ 6) $\times$ 10$^{43}$ ergs s$^{-1}$ = (0.15 $-$
0.40)\% $L_{\rm AGN}$, where $L_{\rm AGN}$ = 1.5 $\times$ 10$^{46}$
ergs s$^{-1}$, derived from the infrared 15-to-30 $\mu$m color and the
prescription of Veilleux et al.\ (2009). These numbers need to be
scaled up by a factor of 5.3 if the CO$-$to$-$H$_2$ conversion factor
is Galactic rather than ULIRG-like, or scaled down by a factor of 2.4
if CO(1$-$0) is optically thin. The results are summarized in Table 4.

\subsection{Comparisons with the {\em Herschel} OH Outflow}

Table 4 compares the mass, momentum, and kinetic energy outflow rates
derived from the new ALMA CO(1$-$0) data cube with the values derived
from the spatially unresolved {\em Herschel} OH 119 $\mu$m spectral
feature (V13; T15). As noted earlier, the measured velocity of the
CO(1$-$0) outflow is remarkably similar to that of the OH outflow
derived from the {\em Herschel} data. Note, however, that the scales
probed by the two data sets are significantly different: modeling of
the {\em Herschel} OH profile suggests a scale for the OH outflow of
$\sim$0.1 -- 1.0 kpc (nominally 300 pc; T15), while the ALMA data show
broad CO line emission possibly extending out to $\sim$ 15 kpc. This
difference in scale between the OH and CO outflows is not unexpected:
OH absorption is produced by gas in front of the source of FIR
continuum, which is compact in ULIRGs, but there is no such
requirement for the detection of the CO line emission. Moreover,
CO(1$-$0) traces the more diffuse low-excitation molecular gas, from
which there may not be excited absorption. This difference in scale is
important since the dynamical parameters of the CO outflow listed in
Table 4 are quantities that are time-averaged over a flow time scale
($R_{\rm out,CO}/V_{\rm out,CO}$) $\sim$7 $\times$ 10$^6$ yrs, while
the published OH-based mass outflow rate is a local
(``instantaneous'') estimate at $R_{\rm out,OH} \sim$ 300 pc, which is
valid for timescales ($\Delta R_{\rm out,OH}/V_{\rm out,OH}$) $\la$
10$^5$ yrs, where $\Delta R_{\rm out,OH}$ $\approx$ 75 pc, the
thickness of outflow shell of molecular material derived from the OH
119 $\mu$m profile. The third row in Table 4 lists the dynamical
quantities time-averaged over the flow time scale $R_{\rm out,OH} /
V_{\rm out,OH}$ = 4 $\times$ 10$^5$ yrs; these quantities are
$R_{\rm out,OH} / \Delta R_{\rm out,OH}$ = 4 times smaller than the
local quantities and comparable to the values derived from the CO
outflow.

Given the well-known short- and long-term variability of F11119$+$3257
(T15) and AGN in general (e.g., Schawinski et al.\ 2010, 2015; Keel et
al.\ 2012a, 2012b, 2015, 2017), it is perhaps surprising to find in
Table 4 that the time-averaged mass, momentum, and energy outflow
rates derived from the CO data are similar to the time-averaged values
derived from the OH data. This suggests that the efficiency of the
quasar at driving the molecular outflow on large scales in
F11119$+$3257 has been relatively stable over the past few $\times$
10$^6$ yrs. We return to this issue in the next section.

\subsection{Comparisons with Published Models}

In T15, we argued that the dynamics of the X-ray wind and OH outflow
were consistent with the models where the OH outflow is an
energy-conserving flow driven by a fast AGN accretion disk wind (e.g.,
Zubovas \& King 2012, 2014; Faucher-Gigu\`ere \& Quataert 2012; Costa
et al.\ 2014; Nims et al.\ 2015). If this is the case, we have by
energy conservation 
\begin{eqnarray}
\dot{P}_{\rm out} & = & f~(V_{\rm wind} / V_{\rm   out})~\dot{P}_{\rm wind} \\
                 & \sim & f~(V_{\rm wind} / V_{\rm out})~(L_{\rm  AGN}/c), 
\end{eqnarray}
 where the quantities with subscript ``out'' refer to the molecular
 outflow, while those with subscript ``wind'' refer to the inner X-ray
 wind. The last equality (eq.\ 5) is valid only if the inner wind is
 radiatively accelerated, i.e.\ $\dot{P}_{\rm wind} \sim L_{\rm AGN} /
 c$, which appears to be the case in F11119$+$3257 (T15; Table 4). The
 efficiency $f$ is defined as the fraction of the kinetic energy in
 the X-ray wind that goes into bulk motion of the swept-up molecular
 material. In T15, an independent estimate of $f$ was derived from the
 ratio of the covering fraction of the OH outflow ($C_{\rm f,out,OH}$)
 to that of the X-ray wind ($C_{\rm f,wind}$). In T15, we derived
 $C_{\rm f,wind} > 0.85$ from the X-ray data and $C_{\rm f,out,OH}$ =
 0.20 $\pm$ 0.05 from the {\em Herschel} data, so $f$ = 0.22 $\pm$
 0.07. In T15, we showed that the above expression for energy
 conservation applies remarkably well to F11119$+$3257, to within the
 (admittedly large) uncertainties of the measurements.

Let us revisit this analysis using the new ALMA data. In the following
discussion, we use the molecular mass, momentum, and energy outflow
rates that are derived assuming a ULIRG-like CO$-$to$-$H$_2$
conversion factor of $\alpha_{\rm CO}$ = 0.8 $M_\odot$ (K km s$^{-1}$
pc$^2$)$^{-1}$. We therefore make the implicit assumption that the
physical state (e.g., density, temperature, metallicity, internal
random/turbulent velocity, external radiation field, etc.) of the
outflowing molecular gas in F11119$+$3257 is similar to that of the
quiescent molecular material in the host ULIRG, from which it
presumably originates. This issue is still a matter of debate,
although the detection of high-density molecular gas entrained in the
outflows of NGC~253 (Walter et al.\ 2017) and other ULIRGs (e.g.,
Mrk~231, Aalto et al.\ 2012, 2015) brings some support to this
assumption.  Given the super-solar metallicity and high total
surface density of ULIRGs (e.g., Rupke et al.\ 2008; Gonz\'alez-Alfonso
et al.\ 2015), we expect $\alpha_{\rm CO}$ to be 2 $-$ 5 $\times$
smaller than the Galactic value (e.g., Bolatto et al.\ 2013a). We also
naively expect a decrease in the CO(1$-$0) optical depth due to the
likelihood of highly turbulent conditions in the emitting gas, but the
detection of high-density molecular gas entrained in the outflow of
NGC~253 favors a high column density (Walter et al\ 2017) and seems to
rule out the conservatively smaller optically thin $\alpha_{\rm CO}$
value of 0.34 $M_\odot$ (K km s$^{-1}$ pc$^2$)$^{-1}$ used by Bolatto
et al.\ (2013a). In the end, we feel that using a ULIRG-like
$\alpha_{\rm CO}$ is the most realistic value for the outflowing
molecular gas, given our current knowledge of the conditions in the
outflowing material, and also a good compromise solution between the
5.3 $\times$ higher values derived assuming Galactic $\alpha_{\rm CO}$
and the 2.4 $\times$ smaller values based on optically thin
$\alpha_{\rm CO}$ (Table 4).

Using the molecular momentum outflow rate based on the ULIRG-like
$\alpha_{\rm CO}$ in eq.\ (4), we derive $f_{\rm CO}$ = 0.02 -- 0.03,
considerably smaller than the value based on the OH outflow using the
local estimates of the energetics ($f_{\rm OH}$ = 0.2; T15), but
comparable to the value we would derive if we use the time-averaged
quantities of Table 4 ($f_{\rm OH}$ = 0.05).  In principle, an
independent value of $f_{\rm CO}$ may be derived from the ratio of the
covering fraction of the CO outflow ($C_{\rm f,out,CO}$) to that of
the X-ray wind ($C_{\rm f,wind} > 0.85$; T15). However, in practice,
the modest angular resolution of our ALMA data, taken in compact-array
configuration, provides only an upper limit on $C_{\rm f,out,CO}$
since the extended emission from the outflowing material seen in
Figure 5 will likely break up into smaller cloudlets when observed at
higher angular resolution (e.g., F08572$+$3915; Jannsen et al.\ 2017,
in prep.), and therefore reduce $C_{\rm f,out,CO}$.  We derive $C_{\rm
  f,out,CO}$ $<$ 0.5, and thus $f_{\rm CO}$ $<$ 0.5, from the
morphology of the high-velocity CO emission on large scale in Figure
5.

Eqs. (4) -- (5) are only valid if the molecular outflow is an
adiabatic energy-driven flow.  However, Table 4 shows that the
molecular momentum outflow rate based on the ULIRG-like $\alpha_{\rm
  CO}$ is only a few times larger than the radiation pressure, $L_{\rm
  AGN}/c$, exerted by the AGN in F11119$+$3257 (Table 4). The
starburst in F11119$+$3257 will contribute an additional term: (1 --
$\alpha_{\rm AGN}$) $L_{\rm BOL}/c$ = [$\frac{(1 - \alpha_{\rm
      AGN})}{\alpha_{\rm AGN}}$] $L_{\rm AGN}/c$ $\sim$ 0.25 $L_{\rm
  AGN}/c$ ($L_{\rm BOL}$ is the bolometric luminosity; Veilleux et
al.\ 2009). A similar statement can be made when considering the
time-averaged OH-based momentum outflow rate.  Thus, the only time we
need to invoke models of energy-conserving flows driven by accretion
disk winds to explain the molecular outflow in F11119$+$3257 is when
we consider the local OH-based momentum outflow rate cited in T15.
The X-ray data of T15 strongly suggest that the accretion disk wind is
momentum-conserving and being driven by radiation pressure from the
AGN. Our data do not allow us to formally rule out the possibility
that the much larger OH and CO outflows are also driven by radiation
pressure, despite the $R^{-2}$ geometric dilution factor of the AGN
radiation field.

It is important to consider the CO and OH outflows together rather
than independently. As discussed in Section 4.2, both are likely
related to one another but refer to significantly different physical
scales ($\sim$0.3 kpc {\em vs} 7 kpc) and time scales ($\sim$4
  $\times$ 10$^5$ yrs {\em vs} $\sim$7 $\times$ 10$^6$ yrs).  The
CO-based momentum outflow rate listed in Table 4 is a quantity that
has been time-averaged over a $\sim$100 $\times$ longer time scale
than the local OH-based momentum outflow rate, so one has to use
caution when making direct comparisons between the two molecular
outflows and with the present properties of the AGN. Indeed, the
CO-based quantities are in much closer agreement with the OH-based
quantities that are time-averaged over the flow time scale ($R / V$).
This general agreement between the energetics of the OH and CO
outflows suggests that the efficiency of the quasar to drive the
large-scale molecular outflow in F11119$+$3257 has remained relatively
constant over the past few $\times$ 10$^6$ yrs. This is not to say
that the luminosity of the AGN in F11119$+$3257 has been constant on
shorter time scales (there is evidence that the hard X-ray flux is
variable on a time scale of a day, although this may be due to
variable absorption columns rather than intrinsic variations;
T15). Our results simply imply that the quasar has not been dormant
for long periods of time over the past few $\times$ 10$^6$ yrs. With
this in mind, it is important to use methods that are insensitive to
short-term AGN variability when estimating the AGN luminosity. Our use
of the global 15-to-30 $\mu$m color (Veilleux et al.\ 2009) to
estimate the fraction of the bolometric luminosity of F11119$+$3257
produced by the AGN, rather than the (variable) hard X-ray luminosity,
mitigates the effects associated with short-term ($\la$ 10$^{3 - 4}$
yrs) AGN variability.

\section{Conclusions}

We report the results of our analysis of deep new ALMA CO(1$-$0) data
on F11119$+$3257 obtained in the compact array configuration
($\sim$2$\farcs$8 resolution). These data are compared with our
findings published in Tombesi et al.\ (2015) of an X-ray detected AGN
accretion disk wind driving a kpc-scale energy-conserving molecular
(OH) outflow in this object. The main results of this analysis are:

\begin{itemize}

\item The CO(1$-$0) spectrum shows the presence of broad wings
  extending $\sim$ $\pm$ 1000 km s$^{-1}$ relative to systemic
  velocity, indicative of a fast CO outflow with velocities comparable
  to those measured from the {\em Herschel} OH 119 $\mu$m line
  profile.

\item Careful photometric and {\em uv}-plane analyses of the ALMA
  data indicate that the broad-wing CO(1$-$0) emission extends on a
  scale of at least $\sim$7 kpc (radius) from the center.  This is the
  largest molecular outflow found so far in a local ULIRG.

\item The mass of molecular gas involved in the CO outflow is (0.6 $-$
  1.4) $\times$ 10$^9$ $M_\odot$, assuming a ULIRG-like $\alpha_{\rm
    CO}$ of 0.8 $M_\odot$ (K km s$^{-1}$ pc$^2$)$^{-1}$. This
  represents $\sim$7-20\% of the quiescent molecular material in the
  host galaxy. The flow time scale ($R_{\rm out,CO}/V_{\rm out,CO}$)
  of this large CO outflow is $\sim$ 7 $\times$ 10$^6$ yrs.  The
  molecular mass, momentum, and energy outflow rates time-averaged
  over the flow time scale are (80 $-$ 200) $M_\odot$ yr$^{-1}$, (6
  $-$ 13) $\times$ 10$^{35}$ dyne = (1.5 $-$ 3.0) $L_{\rm AGN}/c$, and
  (3 $-$ 6) $\times$ 10$^{43}$ ergs s$^{-1}$ = (0.15 $-$ 0.40)\%
  $L_{\rm AGN}$, respectively ($L_{\rm AGN}$ = 1.5 $\times$ 10$^{46}$
  ergs s$^{-1}$ is the AGN luminosity derived from the infrared
  15-to-30 $\mu$m color and the prescription of Veilleux et
  al.\ 2009).

\item At face value, the CO-based momentum outflow rate is not
  inconsistent with the scenario where the CO outflow is
  momentum-conserving and driven by the AGN radiation pressure. This
  is a different picture than that proposed by Tombesi et al.\ (2015),
  who used the local (``instantaneous'') value of the OH-based
  momentum outflow rate estimated at $R \sim$ 300 pc and valid for
  timescales $\Delta R_{\rm out,OH}/R_{\rm out,OH}$ $\la$ 10$^5$ yrs
  i.e.\ nearly two orders of magnitude shorter than the flow time
  scale of the CO outflow ($\Delta R_{\rm out,OH}$ is the thickness of
  the outflowing shell of molecular material).  In contrast, the
  OH-based dynamical quantities time-averaged over the flow time scale
  $R_{\rm out,OH}/V_{\rm out,OH}$ are $R_{\rm out,OH}/\Delta R_{\rm
    out,OH}$ = 4 times smaller than the local quantities and thus
  comparable to the values derived from the CO outflow. These results
  suggest that the efficiency of the quasar to drive the large-scale
  molecular outflow in F11119$+$3257 has remained relatively stable
  over the past few $\times$ 10$^6$ yrs.
\end{itemize}

The modest angular resolution of the ALMA data set is a major
limitation of our analysis. It will be important to revisit
F11119$+$3257 at higher resolution to constrain the morphology (e.g.,
distribution and clumpiness) and velocity field of the CO outflow on
kpc and sub-kpc scales. In the long term, F11119$+$3257 may serve as a
local template for future ALMA OH observations in the distant
Universe, where accretion disk winds are below the detection limits of
current X-ray observatories. The launch in the next few years of the
{\em X-Ray Astronomy Recovery Mission (XARM)}, the replacement to
  {\em Hitomi (ASTRO-H)}, will change the landscape and allow us to
search for X-ray winds in the X-ray brightest ULIRGs with known
molecular outflows as well as some high-redshift
quasars. F11119$+$3257 will be the standard bearer for these future
studies.

\acknowledgements 

We thank the anonymous referee for his/her constructive comments that
improved this paper.  Partial support from the National Science
Foundation through grants AST-1207785 (S.V.), AST-0955836 (A.B.), and
AST-1412419 (A.B.) is gratefully acknowleged. S.V.\ also acknowledges
NASA/ADAP grant NNX16AF24G. F.T.\ acknowledges partial support by NASA
through NuSTAR award NNX15AV21G. Basic research in IR astronomy at NRL
is funded by the US ONR. This paper makes use of the following ALMA
data: ADS/JAO.ALMA \#2015.1.00305.S. ALMA is a partnership of ESO
(representing its member states), NSF (USA), and NINS (Japan),
together with NRC (Canada), and NSC and ASIAA (Taiwan), in cooperation
with the Republic of Chile. The Joint ALMA Observatory is operated by
ESO, AUI/ NRAO and NAOJ. The NRAO is a facility of the National
Science Foundation operated under cooperative agreement by Associated
Universities, Inc. This work also made use of NASA's Astrophysics Data
System Abstract Service and the NASA/IPAC Extragalactic Database
(NED), which is operated by the Jet Propulsion Laboratory, California
Institute of Technology, under contract with the National Aeronautics
and Space Administration.

{\em Facilities: ALMA, Herschel, Suzaku}

\clearpage

\capstartfalse
\begin{deluxetable*}{lcccc}
\tablecolumns{5}
%\tabletypesize{\scriptsize}
\tablecaption{Measured Quantities from Two-Gaussian Fits of the Integrated CO(1$-$0) Emission}
\tablewidth{0pt}
\tablehead{
\colhead{Component} & \colhead{V$_{\rm central}$} & \colhead{FWHM}
& \colhead{Integrated Flux} & \colhead{Peak Flux} \\
\colhead{} & \colhead{(km~s$^{-1}$)} & \colhead{(km~s$^{-1}$)}
& \colhead{(Jy km s$^{-1}$)} & \colhead{(mJy)} \\
\colhead{(1)} & \colhead{(2)} & \colhead{(3)} & \colhead{(4)} &
\colhead{(5)} }
\startdata
%\hline \noalign {\smallskip}
\multicolumn{5}{c}{3\arcsec-radius aperture} \\
\hline \noalign {\smallskip}
Narrow Gaussian & $-$32 $\pm$ 2& 226 $\pm$ 4& 4.63 $\pm$ 0.07& 19.03 $\pm$ 0.25\\
Broad Gaussian & $-$11 $\pm$ 53& 1113 $\pm$ 171& 0.66 $\pm$ 0.07& 0.55 $\pm$ 0.14\\
\hline \noalign {\smallskip}
\multicolumn{5}{c}{5\arcsec$-$radius aperture} \\
\hline \noalign {\smallskip}
Narrow Gaussian & $-$32 $\pm$ 2& 224 $\pm$ 4& 5.14 $\pm$ 0.10& 21.33 $\pm$ 0.25\\
Broad  Gaussian & +47 $\pm$ 48& 1068 $\pm$ 168& 0.98 $\pm$ 0.10& 0.86 $\pm$ 0.18
\enddata
% \tablecomments{}
\end{deluxetable*}

% \capstartfalse
\begin{deluxetable*}{lccccc}
\tablecolumns{6}
%\tabletypesize{\scriptsize}
\tablecaption{Measured Quantities from Residual Map after Removal of the Rotating Disk}
\tablewidth{0pt}
\tablehead{
\colhead{Component} & \colhead{Velocity Range} & \colhead{Integrated Flux} & \colhead{Size (FWHM)$^{(a)}$} & \colhead{RA Offset} & \colhead{DEC Offset}\\
\colhead{} & \colhead{(km s$^{-1}$)} & \colhead{(Jy km s$^{-1}$)} & \colhead{(arcsec)} & \colhead{(arcsec)} & \colhead{(arcsec)}\\
\colhead{(1)} & \colhead{(2)} & \colhead{(3)} & \colhead{(4)} & \colhead{(5)} & \colhead{(6)} }
\startdata
%\hline \noalign {\smallskip}
Blue + red wings & [$-$820, $-$400], [+280, +800]& 0.40 $\pm$ 0.10& (5.1 $\times$ 2.8) $\pm$ 0.1 & +0.22 $\pm$ 0.05 & $-$0.75 $\pm$ 0.10
\enddata
\tablecomments{$^{(a)}$ Not corrected for the beam size (3$\farcs$46 $\times$ 2$\farcs$21 FWHM)}
\end{deluxetable*}

%\capstartfalse
\begin{deluxetable*}{lccccc}
\tablecolumns{6}
%\tabletypesize{\scriptsize}
\tablecaption{Measured Quantities from {\em UV} Plane Fitting}
\tablewidth{0pt}
\tablehead{
\colhead{Component} & \colhead{Velocity Range} & \colhead{Integrated Flux} & \colhead{Size (FWHM)} & \colhead{RA Offset} & \colhead{DEC Offset}\\
\colhead{} & \colhead{(km s$^{-1}$)} & \colhead{(Jy km s$^{-1}$)} & \colhead{(arcsec)} & \colhead{(arcsec)} & \colhead{(arcsec)}\\
\colhead{(1)} & \colhead{(2)} & \colhead{(3)} & \colhead{(4)} & \colhead{(5)} & \colhead{(6)} }
\startdata
%\hline \noalign {\smallskip}
\multicolumn{6}{c}{Fitter: {\em uvmodelfit}} \\
\hline \noalign {\smallskip}
Blue + red wings & [$-$820, $-$380], [+440, +800] & 0.31 $\pm$ 0.05 & 3.9 $\pm$ 1.0 & +0.12 $\pm$ 0.21 & $-$0.44 $\pm$ 0.30\\ 
\hline \noalign {\smallskip}
\multicolumn{6}{c}{Fitter: {\em uvmultifit}} \\
\hline \noalign {\smallskip}
Blue wing & [$-$820, $-$380] & 0.22 $\pm$ 0.08 & 3.8 $\pm$ 2.0 & $-$0.35 $\pm$ 0.5 & $-$0.03 $\pm$ 0.60\\ 
Red wing & [+440, +800] & 0.14 $\pm$ 0.07 & 5.4 $\pm$ 3.5 & +0.2 $\pm$ 0.7 & $-$0.98 $\pm$ 1.20
\enddata
% \tablecomments{}
\end{deluxetable*}

\clearpage

%\capstartfalse
\begin{deluxetable*}{lccccccc}
\tablecolumns{8}
%\tabletypesize{\scriptsize}
\tablecaption{Derived Properties of the Small- and Large-Scale Outflows in 
 F11119$+$3257}
\tablewidth{0pt}
\tablehead{
\colhead{Outflow} & \colhead{Outflow} & \colhead{Radius}
& \colhead{Radius} & \colhead{Covering} & \colhead{$\dot{M}$} &
\colhead{$\dot{P}$} & \colhead{$\dot{E}$} \\
\colhead{Type} & \colhead{Velocity} & \colhead{(lower limit)} & \colhead{(upper limit)} & \colhead{Fraction} & \colhead{[$M_\odot$ yr$^{-1}$]} & \colhead{[$L_{\rm AGN}$/c]} & \colhead{[$L_{\rm AGN}$]} \\
\colhead{(1)} & \colhead{(2)} & \colhead{(3)} & \colhead{(4)} &
\colhead{(5)} & \colhead{(6)} & \colhead{(7)} & \colhead{(8)}
}
\startdata
%\hline \noalign {\smallskip}
Accretion Disk Wind$^{(a)}$ & 0.255 $\pm$ 0.011 c    &15$r_s$ &900$r_s$   &$>$0.85  & 1.5 $-$ 4.5 $^{(b)}$ & 0.4 $-$ 3.0 $^{(c)}$ & (6 $-$ 50)\% $^{(d)}$ \\
OH Outflow (local)$^{(e)}$ & 1000 $\pm$ 200 km s$^{-1}$&0.1 kpc &1.0 kpc & 0.20$\pm$0.05&250 $-$ 2000 $^{(f)}$ & 3.5 $-$ 25 $^{(g)}$ & (0.5 $-$ 5.0)\% $^{(h)}$ \\
{\bf OH Outflow (average)$^{(i)}$} & {\bf 1000 $\pm$ 200 km s$^{-1}$}&{\bf 0.1 kpc} &{\bf 1.0 kpc} & {\bf 0.20$\pm$0.05}&{\bf 60 $-$ 500 $^{(j)}$} & {\bf 1.0 $-$ 6 $^{(g)}$} & {\bf (0.1 $-$ 1.0)\% $^{(h)}$ }\\
{\bf CO Outflow (ULIRG-like)$^{(k)}$} & {\bf 1000 $\pm$ 200 km s$^{-1}$}&{\bf $<$4.0 kpc} &{\bf 15 kpc} & {\bf $<$0.50$^{(l)}$}&{\bf 80 $-$ 200 $^{(m)}$} & {\bf 1.5 $-$ 3 $^{(n)}$} & {\bf (0.15 $-$ 0.40)\% $^{(o)}$  }\\
CO Outflow (Galactic)$^{(p)}$ & 1000 $\pm$ 200 km s$^{-1}$&$<$4.0 kpc &15 kpc & $<$0.50$^{(l)}$&400 $-$ 1000 $^{(m)}$ &8 $-$ 16 $^{(n)}$ & (0.80 $-$ 2.0)\% $^{(o)}$ \\
CO Outflow (optically thin)$^{(q)}$ & 1000 $\pm$ 200 km s$^{-1}$&$<$4.0 kpc &15 kpc & $<$0.50$^{(l)}$&30 $-$ 90 $^{(m)}$ &0.6 $-$ 1.3 $^{(n)}$ & (0.06 $-$ 0.17)\% $^{(o)}$ 
\enddata
  
\tablecomments{Boldfaced entries indicate favored estimates. Column
  (1): This table lists the physical properties of three different
  outflows: (i) the X-ray wind on the scale of the accretion disk
  first reported in T15, (ii) the {\em Herschel}-detected OH outflow
  first reported in V13, and (iii) the ALMA-detected CO outflow
  reported in the present paper. Column (2): Estimate of the outflow
  velocity. Column (3): Lower limit on the size of the outflow.
  Column (4): Upper limit on the size of the outflow. Column (5):
  Estimate of the fraction of the sky covered by the outflowing
  material. Column (6): Mass outflow rate in $M_\odot$
  yr$^{-1}$. Column (7): Momentum flux normalized to the radiation
  pressure, $L_{\rm AGN}$/$c$. Column (8): Mechanical power normalized
  to the AGN luminosity, $L_{\rm AGN}$ = 1.5 $\times$ 10$^{46}$ ergs
  s$^{-1}$, derived from the infrared 15-to-30 $\mu$m color and the
  prescription of Veilleux et al.\ (2009).\\ \\ $^{(a)}$ Quantities
  derived from the {\em Suzaku} data of T15. $^{(b)}$ $\dot{M}_{\rm
    wind}$ = 1.5 $\times$ ($r_{\rm wind}$/15 $r_s$) ($N_H$/6.4
  $\times$ 10$^{24}$ cm$^{-2}$) $\times$ ($C_{\rm f,wind}$/1.0)
  $\times$ ($V_{\rm wind}$/0.255c) $M_\odot$ yr$^{-1}$, where $r_{\rm
    wind}$ is the wind radius, $r_s$ is the Schwarzschild radius of
  the SMBH in F11119$+$3257 ($M_{\rm BH} = 1.6 \times 10^7$ $M_\odot$;
  Kawakatu et al.\ 2007), $N_H$ is column density of the (fully
  ionized) wind, and $C_{\rm f,wind}$ is the wind covering fraction.
  $^{(c)}$ $\dot{P}_{\rm wind}$ = $\dot{M}_{\rm wind}$ $\times$
  $V_{\rm wind}$. $^{(d)}$ $\dot{E}_{\rm wind}$ = $\frac{1}{2}$
  $\dot{M}_{\rm wind}$ $\times$ $V_{\rm wind}^2$. $^{(e)}$ Local
  (``instantaneous'') quantities derived by T15 from the {\em
    Herschel} OH 119 $\mu$m presented in V13. $^{(f)}$ $\dot{M}_{\rm
    out,OH}$ = 800 $\times$ ($R_{\rm out,OH}$/300 pc)$^2$ $\times$
  ($n_H$/100 cm$^{-3}$) $\times$ ($C_{\rm f,out,OH}$/0.2) $\times$
  ($V_{\rm out,OH}$/1000 km s$^{-1}$) $M_\odot$ yr$^{-1}$ = $M_{\rm
    out,OH}$ $V_{\rm out,OH}$ $\Delta R_{\rm out,OH}^{-1}$, where
  $R_{\rm out,OH}$ is the radius of the OH outflow, $n_H$ is the
  Hydrogen number density, $C_{\rm f,out,OH}$ is the OH outflow
  covering fraction, $M_{\rm out,OH}$ is the total outflowing mass of
  molecular gas, and $\Delta R_{\rm out,OH}$ is the thickness of the
  outflowing shell (= 75 pc). $^{(g)}$ $\dot{P}_{\rm out,OH}$ =
  $\dot{M}_{\rm out,OH}$ $\times$ $V_{\rm out,OH}$.  $^{(h)}$
  $\dot{E}_{\rm out,OH}$ = $\frac{1}{2}$ $\dot{M}_{\rm out,OH}$
  $\times$ $V_{\rm out,OH}^2$.  $^{(i)}$ Time-averaged quantities
  derived from the {\em Herschel} OH 119 $\mu$m presented in V13.
  $^{(j)}$ $\dot{M}_{\rm out,OH}$ = 200 $\times$ ($R_{\rm out,OH}$/300
  pc) $\times$ ($N_H$/2.3 $\times$ 10$^{22}$ cm$^{-2}$) $\times$
  ($C_{\rm f,out,OH}$/0.2) $\times$ ($V_{\rm out,OH}$/1000 km
  s$^{-1}$) $M_\odot$ yr$^{-1}$ = $M_{\rm out,OH}$ $V_{\rm out,OH}$
  $R_{\rm out,OH}^{-1}$, where $R_{\rm out,OH}$ is the radius of the
  OH outflow, $N_H$ is the Hydrogen column density, $C_{\rm f,out,OH}$
  is the OH outflow covering fraction, $M_{\rm out,OH}$ is the total
  outflowing mass of molecular gas.  $^{(k)}$ Quantities derived from
  the ALMA CO(1$-$0) data using a ULIRG-like $\alpha_{\rm CO}$ of 0.8
  $M_\odot$ (K km s$^{-1}$ pc$^2$)$^{-1}$. $^{(l)}$ The covering
  fraction of the CO outflow is estimated from the patchiness of the
  high-velocity CO emission on large scale in Figure 5 (see Section
  4.3 for more detail). $^{(m)}$ $\dot{M}_{\rm out,CO}$ = 140
  (${M}_{\rm out,CO}$/1 $\times$ 10$^9$ $M_\odot$) $\times$ ($R_{\rm
    out,CO}$/7 kpc) ($V_{\rm out,CO}$/1000 km s$^{-1}$)$^{-1}$
  $M_\odot$ yr$^{-1}$. $^{(n)}$ $\dot{P}_{\rm out,CO}$ = $\dot{M}_{\rm
    out,CO}$ $\times$ $V_{\rm out,CO}$.  $^{(o)}$ $\dot{E}_{\rm
    out,CO}$ = $\frac{1}{2}$ $\dot{M}_{\rm out,CO}$ $\times$ $V_{\rm
    out,CO}^2$.  $^{(p)}$ Quantities derived from the ALMA CO(1$-$0)
  data using a Galactic $\alpha_{\rm CO}$ = 4.3 $M_\odot$ (K km
  s$^{-1}$ pc$^2$)$^{-1}$.  $^{(q)}$ Quantities derived from the ALMA
  CO(1$-$0) data using an optically thin $\alpha_{\rm CO}$ = 0.34
  $M_\odot$ (K km s$^{-1}$ pc$^2$)$^{-1}$. }
  
\end{deluxetable*}

\capstarttrue

\clearpage

\begin{figure*}
\epsscale{0.8}
\centering
\includegraphics[width=1.0\textwidth,angle=0]{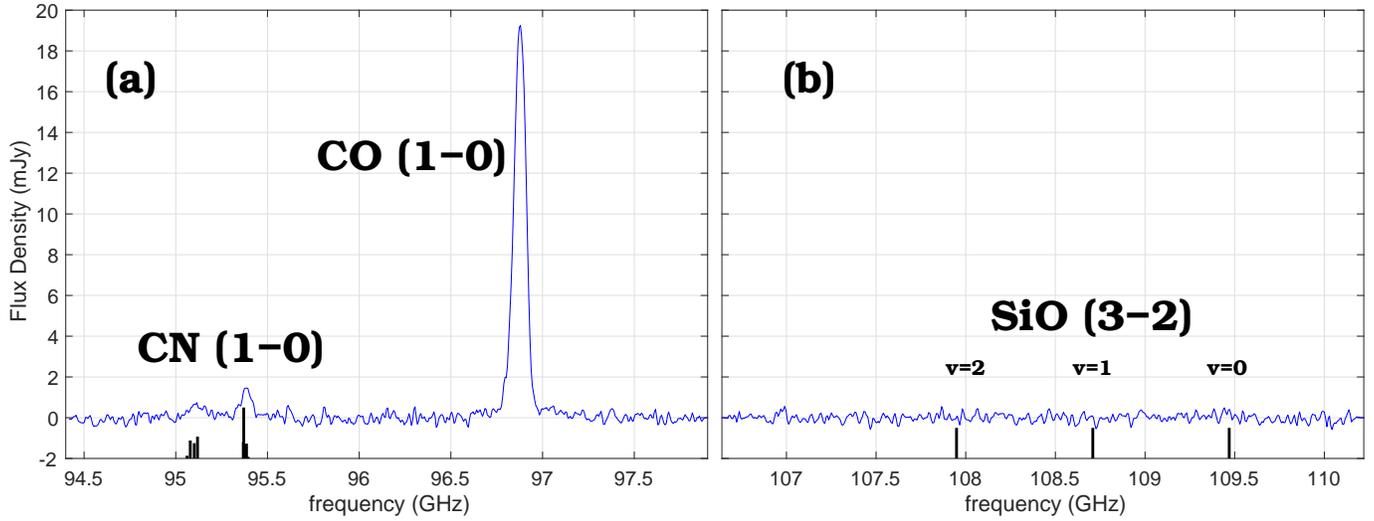}
\caption{Full continuum-subtracted USB spectrum integrated inside a
  3$\arcsec$-radius circular aperture: (a) 94.4 $-$ 97.9 GHz, (b)
  106.6 $-$ 110.2 GHz. Channels are 20 km~s$^{-1}$ wide but Hanning
  velocity smoothing was carried out to result in a spectral
  resolution of $\sim$40 km~s$^{-1}$.  The vertical black lines in (a)
  show the expected positions for the CN (1-0) hyperfine components,
  with the relative intensities observed in Orion (Turner \& Gammon
  1975). The SiO $v$ = 0 (3-2), $v$ = 1 (3-2), and $v$ = 2 (3-2)
  transitions are not detected in (b).}
\end{figure*}

\clearpage

\begin{figure*}
\epsscale{0.8}
\centering
\includegraphics[width=1.0\textwidth,angle=0]{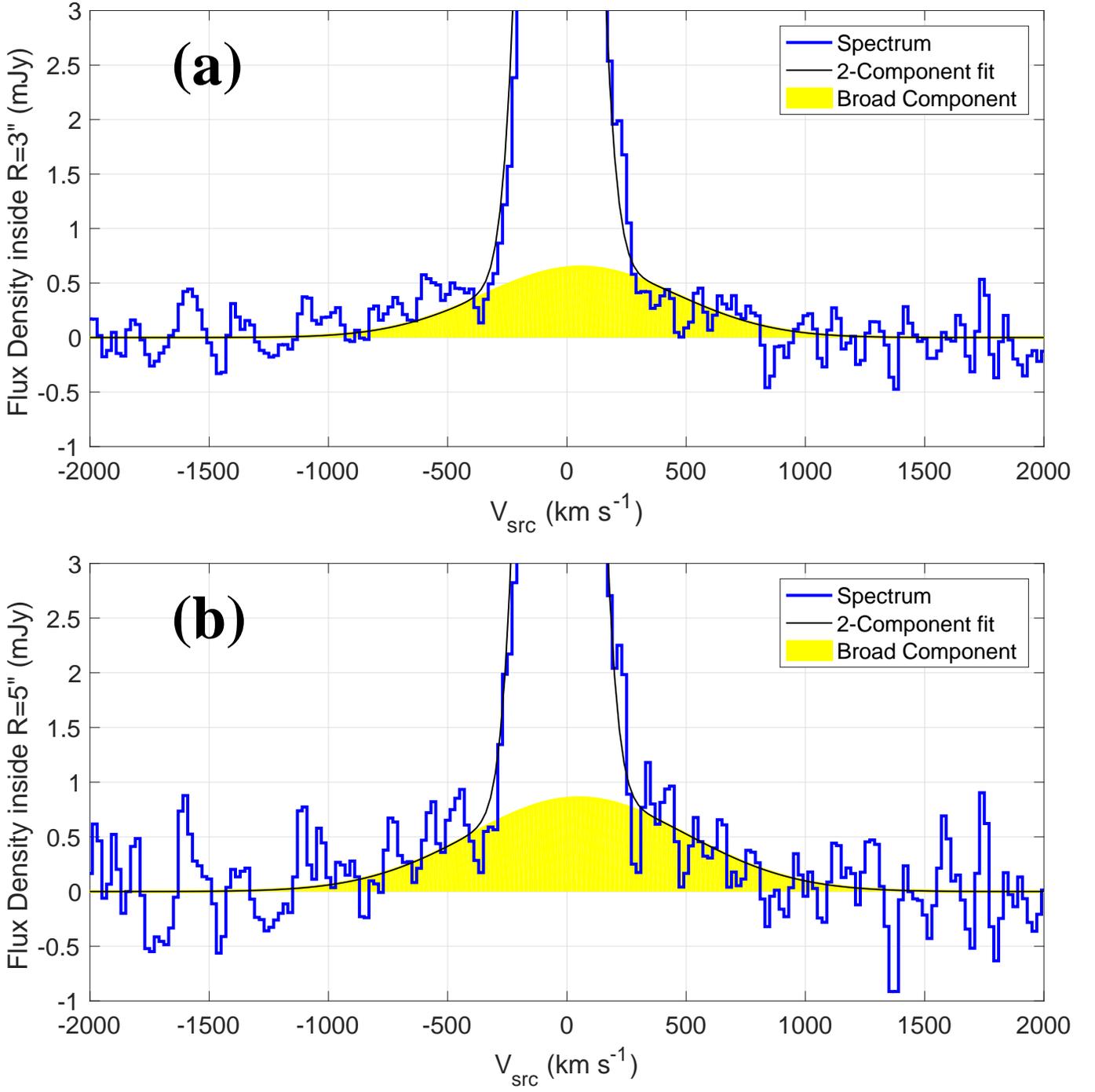}
\caption{Simultaneous two-Gaussian fit to the CO(1$-$0) line emission
  within (a) a 3\arcsec-radius circular aperture centered on
  F11119$+$3257, and (b) a 5\arcsec-radius circular aperture centered
  on F11119$+$3257. The CO fluxes in the broad and narrow components
  are listed in Table 1.}
\end{figure*}

\clearpage

\begin{figure*}
\epsscale{0.8}
\centering
\includegraphics[width=1.0\textwidth,angle=0]{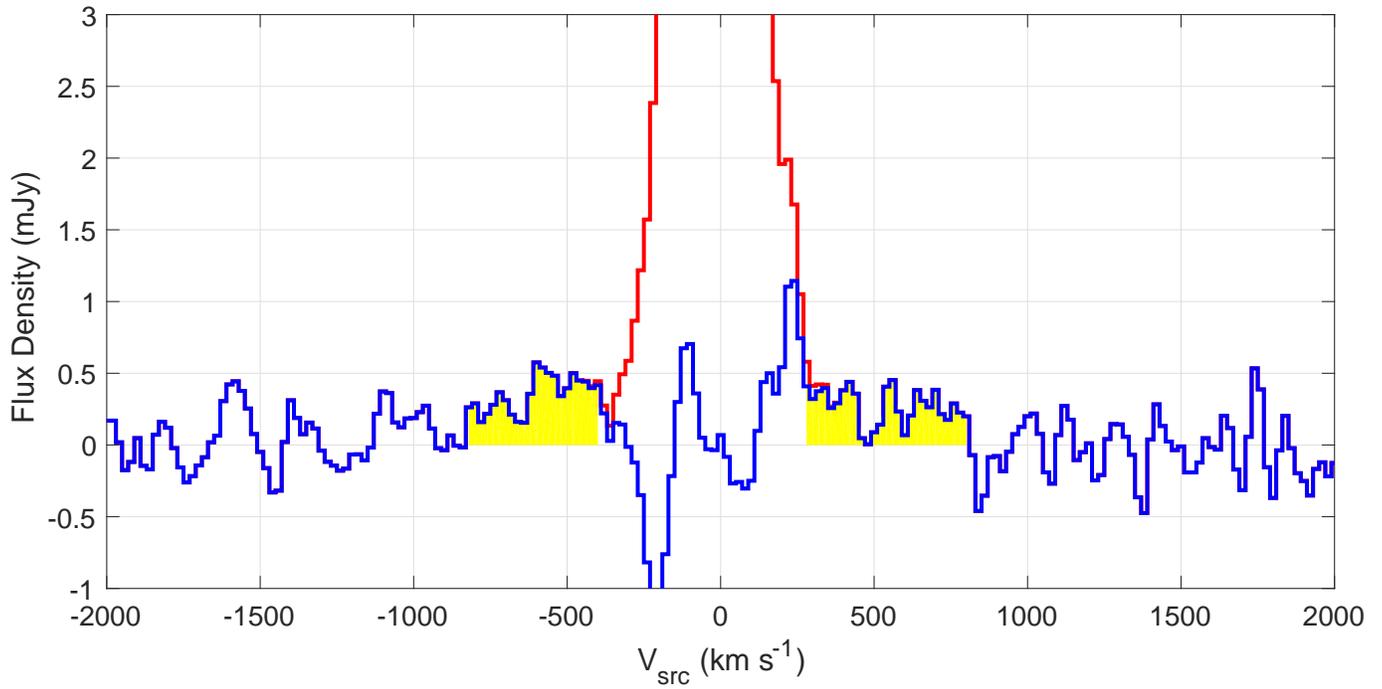}
\caption{Same as Figure 2a, but here the blue line shows the residuals
  after fitting and removing a two-dimensional Gaussian source model
  to each 20 km~s$^{-1}$ channel, representative of the gas in pure
  rotation. See text in Section 2 for more detail on the removal
  method. The yellow region shows the high-velocity emission in CO,
  from $-$820 to $-$400 km s$^{-1}$ and $+$280 to $+$800 km
    s$^{-1}$, which cannot be accounted for by the gas in pure
  rotation. }
\end{figure*}

\clearpage

\begin{figure*}
\epsscale{0.8}
\centering
\includegraphics[width=1.0\textwidth,angle=0]{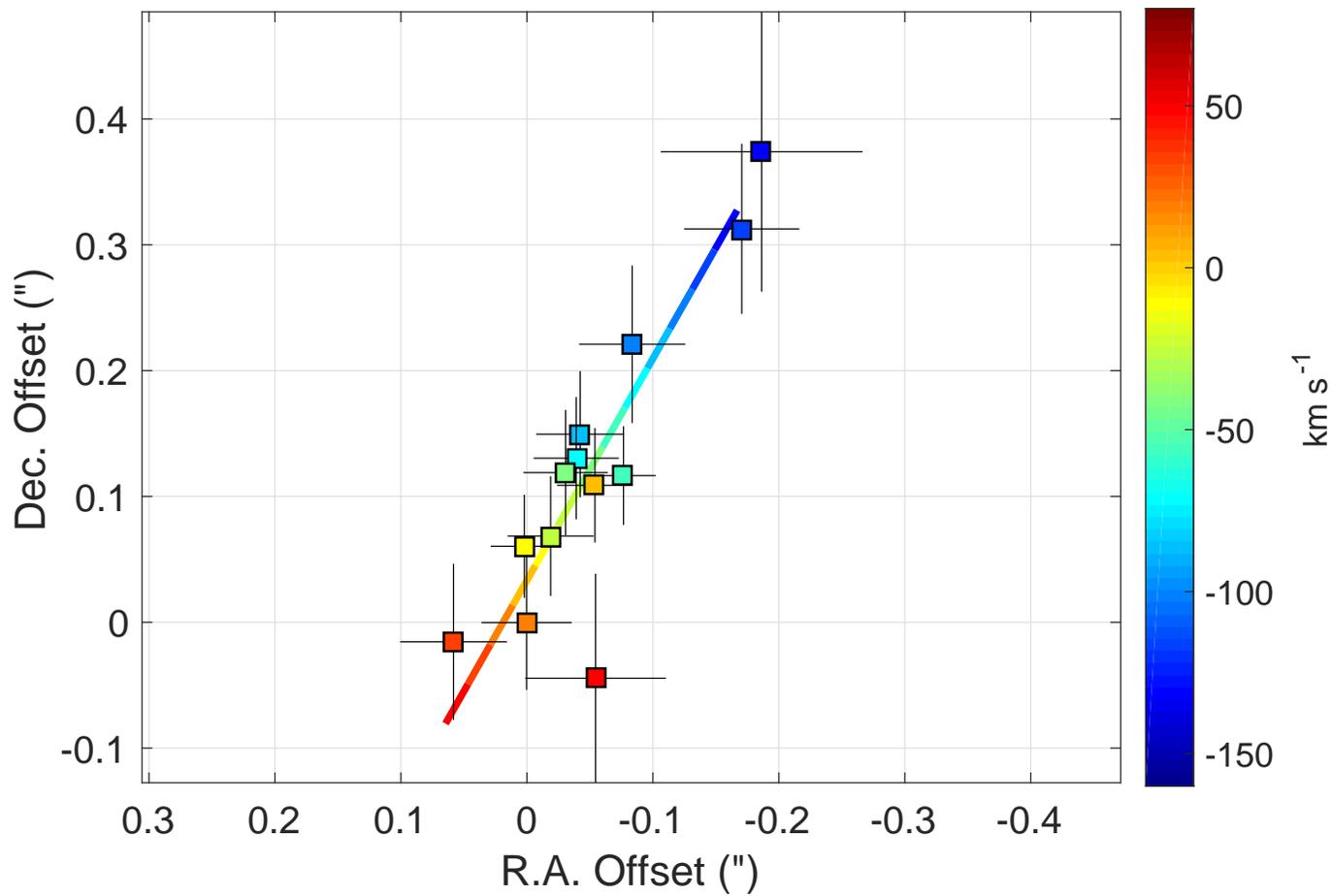}
\caption{Velocity gradient due to rotation in the host galaxy derived
  from the ALMA CO (1 $-$ 0) line emission. The linear scale is 3.19
  kpc per arcsecond. See Section 2 for more detail on the
  derivation. }
\end{figure*}

\clearpage

\begin{turnpage}

\begin{figure*}
\centering
\includegraphics[width=1.3\textwidth,angle=0]{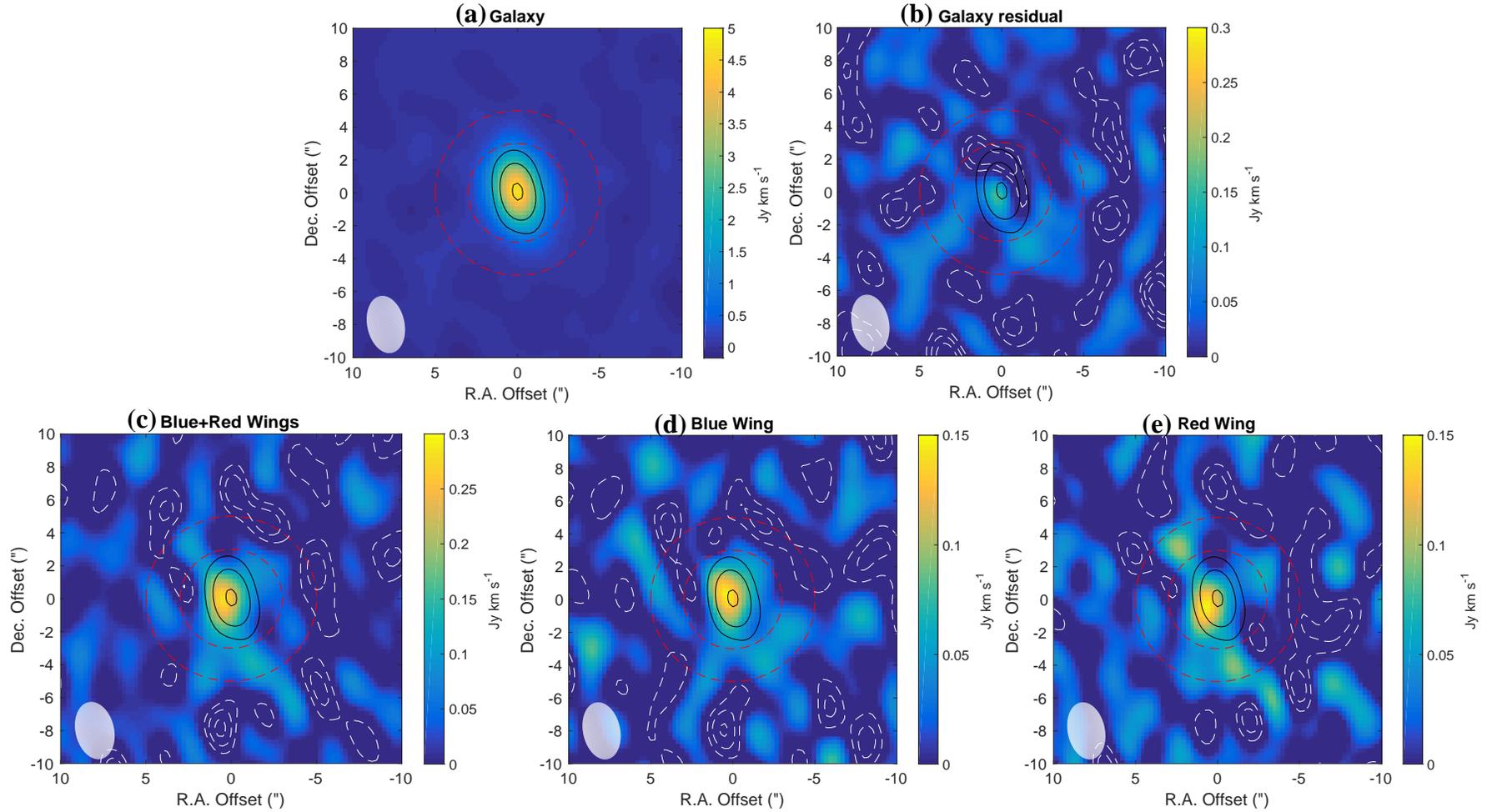}
\caption{Maps of the CO(1$-$0) emission from the various
    kinematic components of F11119$+$3257: (a) the rotating disk, (b)
    ``residuals'' channels between $-$400 and $+$280 km s$^{-1}$ after
    subtraction of the rotating material, (c) blue + red wings,
    i.e.\ the ``residual'' channels between $-$820 and $-$380 km
    s$^{-1}$ and between $+$280 and $+$800 km s$^{-1}$, (d) blue wing
    only, i.e.\ between $-$820 and $-$380 km s$^{-1}$, (e) red wing
    only, i.e.\ between $+$280 and $+$800 km s$^{-1}$.  The linear
    scale is 3.19 kpc per arcsecond. For each panel, the color scale
    on the right indicates the flux level (note that the panels are on
    different scales). The white contours indicate $-$1, $-$2, and
    $-$3 $\times$ the rms noise (= 0.033, 0.033, 0.04, 0.026, and
    0.029 Jy km s$^{-1}$ for panels a, b, c, d, and e, respectively).
  The black contours show the USB continuum emission (0.1, 0.25, and
  0.5 mJy). The beam size is shown in the lower left corner of each
  panel, and the 3$\arcsec$- and 5$\arcsec$-radius circular apertures
  centered on the CO peak are shown as red dashed circles. Note that
  the emission from the high-velocity gas is extended and offset from
  the continuum emission and the rotating disk.}
\end{figure*}

\end{turnpage}

\clearpage

\begin{figure*}
\epsscale{0.8}
\centering
\includegraphics[width=1.0\textwidth,angle=0]{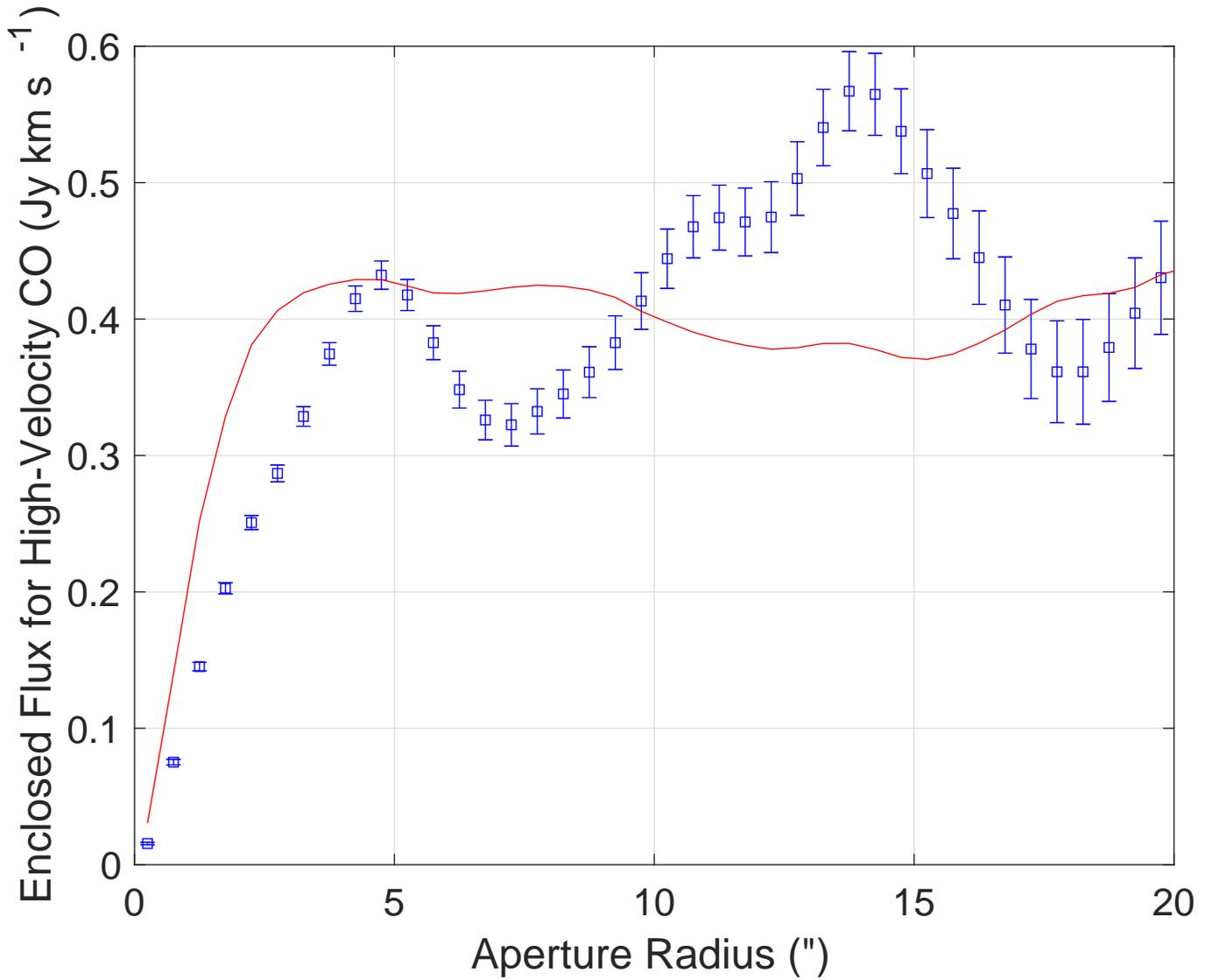}
\caption{Aperture photometry on the high-velocity CO (1 $-$ 0)
  emission shown in Figures 3 and 5. The integrated high-velocity CO
  line flux is plotted in blue as a function of the radius of the
  circular aperture. For comparison, the integrated continuum flux,
  which is unresolved, is shown in red. The flux peaks around $R$ = (5
  $\pm$ 1.5)$\arcsec$ and then stays around $\sim$0.4 $\pm$ 0.1 Jy km
  s$^{-1}$. A radius of (5 $\pm$ 1.5)$\arcsec$ measured on the image
  corresponds to an actual radius of 4$\farcs$8$^{+1.5}_{-2.7}$ =
  15$^{+5}_{-8}$ kpc after correcting for the beam size -- this is our
  best estimate of the maximum extent of the CO outflow.}
\end{figure*}

\clearpage

\end{document}